\documentclass[12pt,letterpaper]{article}

\usepackage{graphicx}
\usepackage{rangecite}
\usepackage{ifpdf}
\usepackage{amsmath}

\def\be{\begin{equation}}
\def\ee{\end{equation}}
\def\bear{\begin{eqnarray}}
\def\eear{\end{eqnarray}}

\def\half{{{1\over 2}}}

\begin{document}


\begin{titlepage}
\begin{flushright} NSF-KITP-11-182 \end{flushright}
\vskip 1in
\begin{center}
{\Large
{Large Representation Recurrences in Large N Random Unitary  
\\ \vskip 0.07in   
Matrix Models}}
\vskip 0.5in
{Joanna L. Karczmarek$^a$ and Gordon W. Semenoff$^{\,a,b}$}
\vskip 0.3in
$^a$~{\it
Department of Physics and Astronomy\\
University of British Columbia
Vancouver, Canada V6T 1Z1}\vskip 0.3in
$^b$~{\it Kavli Institute for Theoretical Physics, University of
California,\\ Santa Barbara, California 93106-4030}
\end{center}
\vskip 0.5in
\begin{abstract}
In a random unitary matrix model at large $N$, we study the properties
of the expectation value of the character of the unitary matrix
in the rank $k$ symmetric tensor representation. We address the
problem of whether the standard semiclassical technique for solving the model in
the large $N$ limit can be applied when the representation is very large,
with $k$ of order $N$. We find that the eigenvalues do indeed
localize on an extremum of the effective potential; however,
for finite but sufficiently large $k/N$,
it is not possible to replace the discrete eigenvalue density with a
continuous one.  Nonetheless, the expectation
value of the character has a well-defined large $N$ limit,
and when the discreteness of the eigenvalues is properly accounted for,
it shows an intriguing approximate periodicity as a function of $k/N$.
\end{abstract}
\end{titlepage}



\section{Introduction and Results}
\label{sec:into}

Unitary matrix models have many applications as a tool to study quantum systems, as well as
an interesting mathematical structure \cite{Morozov:2009jv}.  They make an
appearance in the two-dimensional lattice Yang-Mills theory that was solved in
the large N limit by Gross and Witten \cite{Gross:1980he}  and they have a long
history of applications to mean field models of lattice gauge theory
\cite{Khokhlachev:1979tx,Makeenko:1981qv,Nicolis:2010yx}
and lattice models with induced QCD
\cite{Kazakov:1992ym,Kogan:1992sn,Kogan:1992pa,Kogan:1992ij,Bergere:2008da,Dobroliubov:1993wi}. 
They have interesting deformations which are related to integrable models
\cite{Brezin:1980rk,Brower:1981vt,Anagnostopoulos:1991ss,Gross:1991ji,Mironov:1994mv,Morozov:1995pb}
 and they have
recently been of interest as effective field theories for the low energy degrees of freedom
in Yang-Mills theory defined on certain compact spaces \cite{Aharony:2003sx,Aharony:2005bq}.
In the latter context, with the appropriate effective action, the matrix integral that we shall be
interested in, shown in equation (\ref{loop}) below, computes
the Polyakov loop expectation value in an effective theory corresponding to finite temperature,
large $N$, four dimensional Yang-Mills theory where the three
dimensional space is a sphere \cite{Grignani:2009ua,Semenoff:2009zz}.

In this Paper, we wish to point out an interesting behavior of the expectation values of
the character of the unitary matrix for representations with center charge of order N,
in the large N limit. We focus on the completely symmetric representations, denoted $R={\cal S}_k$,
corresponding to a Young tableaux
with a single row of $k$ boxes.  We will take $k$ large, $k\sim N$, and typically in the region
where $k$ is somewhat larger than $N$.  We strongly suspect that the behavior we find also
occurs for other representations, but we shall not analyze them here.\footnote{
Completely antisymmetric representations whose Young tableaux is a single
  column can also readily be analyzed and have some interesting phase
  structure \cite{Karczmarek:2010ec},  but due to the limit of $N$ boxes
  on the length of a column in their Young Tableau, they do not display the behavior that we discuss here.}
The matrix integrals that we study have the generic form
\be
\langle \mathrm{Tr}_R U \rangle =
{\int [dU] e^{-S[U]} \mathrm{Tr}_R U
\over \int [dU] e^{-S[U]} }~,
\label{loop}
\ee
where $U$ is an $N\times N$ unitary matrix, $[dU]$ is the Haar measure for unitary matrices,
$S[U]$ is a class function of $U$ (in that it obeys
$S[U]=S[VUV^\dagger]$ for any unitary matrix $V$) and
$R$ is an irreducible representation of $SU(N)$. $S[U]$ is of order $N^2$ in the sense
that $S [{\cal I}] \sim N^2$ where ${\cal I}$ is the unit matrix, when $N$ is large.

For later reference, define $\Gamma_k$ by the relation
\be
\langle \mathrm{Tr}_{{\cal S}_k} U \rangle = e^{-N \Gamma_k}~.
\ee

In the following, we shall consider a unitary matrix model with a generic action
equipped with the properties that we have listed above, although in many instances,
to be concrete, we will use the Gross-Witten model (\ref{grosswitten}) since its
eigenvalue density is known explicitly.
In the Gross-Witten model \cite{Gross:1980he},
\be\label{grosswitten}
S[U]=- \frac{N}{\lambda}\left[{\rm Tr}~U+ {\rm Tr}~U^\dagger\right]~.
\ee
The matrix model in (\ref{loop}) is what is referred to as an ``eigenvalue model''
since it is possible to use the symmetry of the action to write the integral
as an integral over diagonal matrices.
Because it is a class function, $S[U]$ is a function of
the eigenvalues only, as is $\mathrm{Tr}_R U$.  Therefore, the above
matrix integral can be written in the standard way as an integral over
the $N$ eigenvalues of the matrix $U$, $\{e^{i\phi_1}, \ldots, e^{i\phi_N}\}$,
\be
\langle \mathrm{Tr}_R U \rangle =
{\int \prod \phi_i ~|\Delta(e^{i\phi_i})|^2~
e^{-S[\phi_i]} ~\mathrm{Tr}_R U
\over \int \prod \phi_i ~|\Delta(e^{i\phi_i})|^2~
e^{-S[\phi_i]}}~,
\label{loop2}
\ee
where $|\Delta(e^{i\phi_i})|^2 = \prod_{i<j} |e^{i\phi_i} - e^{i\phi_j}|^2$
is the Vandermonde determinant.
To evaluate this expression at large $N$, several assumptions
are usually made:
\begin{enumerate}
\item {\bf Saddle point approximation}
First there is the assumption that there is a good large N limit where the
integral can be evaluated in a saddle point approximation.   This includes
the assumption that the effective potential
$S_{\rm eff}(\phi_i)=S[\phi_i] - \ln |\Delta(e^{i\phi_i})|^2$ has
isolated minima and the magnitude of the potential itself for generic values of $\phi_i$
is of order $N^2$. Then, for small enough representations $R$,
the expectation value in equation (\ref{loop2}) can be computed
by the saddle point approximation.  We will denote the location of the
minimum of $S_{\rm eff}(\phi_i)$ by $\hat \phi_j$ and assume for simplicity that
$-\pi\leq\hat \phi_1 \leq \hat \phi_2 \leq \ldots \leq \hat \phi_N\leq\pi$.
With  $\hat U =
\mathrm{diag}(e^{i\hat\phi_1}, \ldots, e^{i\hat\phi_N})$, we have simply
\be \label{saddle-point}
\ln \langle \mathrm{Tr}_R U \rangle ~=~ \ln \mathrm{Tr}_R \hat U~+~{\cal O}(1/N^2)~.
\ee
\item {\bf Probe approximation}
The saddle point approximation can still be applied to the integral at large $N$
where the representation $R$ is large, corresponding to a Young Tableau with $k$ boxes and $k$ is
of order $N$, that is, $\frac{k}{N}$ remains finite as $N\to\infty$ \cite{Hartnoll:2006is,Yamaguchi:2006tq,Okuyama:2006jc,Semenoff:2009zz}.
 Moreover,  to compute the leading term, which is of order $N$ in $\ln \langle \mathrm{Tr}_R U \rangle $,
  it is sufficient to assume that the position of the
saddle point is not affected by the presence of an operator $\mathrm{Tr}_R U$ in the integrand.
 The upshot is that,  for these large representations, (\ref{saddle-point}) is replaced
by
\be \label{probe0}
\ln \langle \mathrm{Tr}_R U \rangle ~=~ \ln \mathrm{Tr}_R \hat U~+~{\cal O}(1 )~.
\ee
and has nontrivial content since $ \ln \langle \mathrm{Tr}_R U \rangle$ is of order $N$.
\item {\bf Continuum approximation}
An essential tool which is used to subsequently evaluate the matrix elements is the distribution function
for the eigenvalues $\hat \phi_i$.   For large $N$, sums over eigenvalues are replaced by
integrals over an eigenvalue density,
\be\label{eigenvaluedensity}
\rho(\phi)=\frac{1}{N}\sum_i\delta(\phi-\hat\phi_i)
\ee
which is assumed to approach a piece-wise smooth function of $\phi$ in the limit where $N$ is large.
It is then employed to compute traces,
for example, \be \mathrm{Tr}~\hat U^k =
\sum_j e^{ik\hat \phi_j} \approx N \int d\phi \rho(\phi) e^{ik\phi}~.\ee
\end{enumerate}

In the following, we shall examine the reliability of the large $N$ expansion for computing expectation values
of the forms $\langle {\rm Tr}~U^k\rangle$ and $\langle {\rm Tr}_{{\cal S}_k}U\rangle$ when $k\sim N$.
We shall see that, for good reasons, the saddle point computation of $\langle {\rm Tr}~U^k\rangle$ fails
once $k \over N$ is large enough.  On the other hand, the expectation value of the character,
$\langle {\rm Tr}_{{\cal S}_k}U\rangle$, is much better behaved.  For it, we shall
find that points 1 and 2, which are essentially the saddle point and probe approximations, hold up to
the largest values of $\frac{k}{N}$ that we can study, and that  $\langle {\rm Tr}_{{\cal S}_k}U\rangle$ has
a well-defined large $N$ limit in the regime where $k \over N$ is of order one.  However, the assumption of a
continuous eigenvalue distribution fails, leading to intriguing behaviour.

To illustrate the latter point, let us assume that all aspects of the large $N$ expansion are valid
and use them to compute $\langle \mathrm{Tr}_{{\cal S}_k}  U\rangle$ in the large $N$ limit.
In that limit, we assume that the eigenvalues are classical, given by those values of $\phi_i$ which minimize
 the effective potential, which we denote by $\hat\phi_i$, and we can simply evaluate
the expectation value by substitution: $\langle \mathrm{Tr}_{{\cal S}_k} U \rangle =  \mathrm{Tr}_{{\cal S}_k}  \hat U $
where $\hat U$ is the classical diagonal matrix.  For concreteness,
let us consider the strong coupling phase of the Gross-Witten model
which has action $S[U]$ given in (\ref{grosswitten}).
It is known to be solved by the eigenvalue density
\be
\rho(\phi) = {1\over 2\pi} \left (1 + 2p \cos(\phi)\right)~,
\ee
where $p =\frac{1}{\lambda} $ when $\lambda\geq 2$.
For the sake of this argument, we will assume that this eigenvalue density can 
be used to compute $\mathrm {Tr}~\hat U^k$. 
Using this density,  traces are given by
\be
\mathrm {Tr}~\hat U^k = \sum_{j=1}^N e^{ik\hat\phi_j} \approx
\left\{ \begin{array}{ll} N & \mathrm{if~} k=0 ~,\\
N  p & \mathrm{if~} |k| =1 ~,\\ 0 & \mathrm{if~} |k| > 1~.\end{array}\right .
\label{grosswittentraces}\ee
The trace of a matrix in the irrep ${\cal S}_k$ can be written in terms of the eigenvalues
as
\be
\mathrm{Tr}_{{\cal S}_k} \hat U = \sum_{j_1 \leq \ldots \leq j_k}  \exp \left (
\sum_{a=1}^k i\hat\phi_{j_a} \right )~,
\label{s}
\ee
which is equivalent, due to the Frobenius formula, to
\be
\label{frobenius}
\mathrm{Tr}_{{\cal S}_k} \hat U  = {1\over k!}\sum_{\sigma \in S_k}
\left ( \prod_{j=1}^p \mathrm{Tr}~\hat U^{l_j} \right )~,
\ee
where $S_k$ is the symmetric group, the sum is over all its elements $\sigma$, 
and where $l_1$,  $l_2$, $\ldots$, $l_p$
are the lengths of the cycles in the decomposition into cycles of the element $\sigma$.

We can now combine equations (\ref{grosswittentraces}) and (\ref{frobenius}) to obtain
\be \mathrm{Tr}_{{\cal S}_k} \hat U  = {1\over k!} (Np)^k \approx
((N/k)pe)^k\ee or
\be  \mathrm{Tr}_{{\cal S}_k} \hat U   \approx e^{- k \ln\left[\frac{k}{N}\frac{1}{ep}\right]}~.
\label{saddle-point-strong-coupling}
\ee

Several things are interesting about this expression:
as expected, since $k\sim N$, $\ln \mathrm{Tr}_{{\cal S}_k} \hat U $ is of order $N$.  For sufficiently
large $k/N$, $\frac{k}{N}>ep$, it is negative. This crossing from a scenario where
$ \mathrm{Tr}_{{\cal S}_k} \hat U $ is an effectively infinite exponential of a
positive quantity of order $N$ to one which is exponentially small in $N$ was interpreted
as a phase transition in references \cite{Grignani:2009ua,Semenoff:2009zz}.

\begin{figure}
\ifpdf
\includegraphics[scale=0.75]{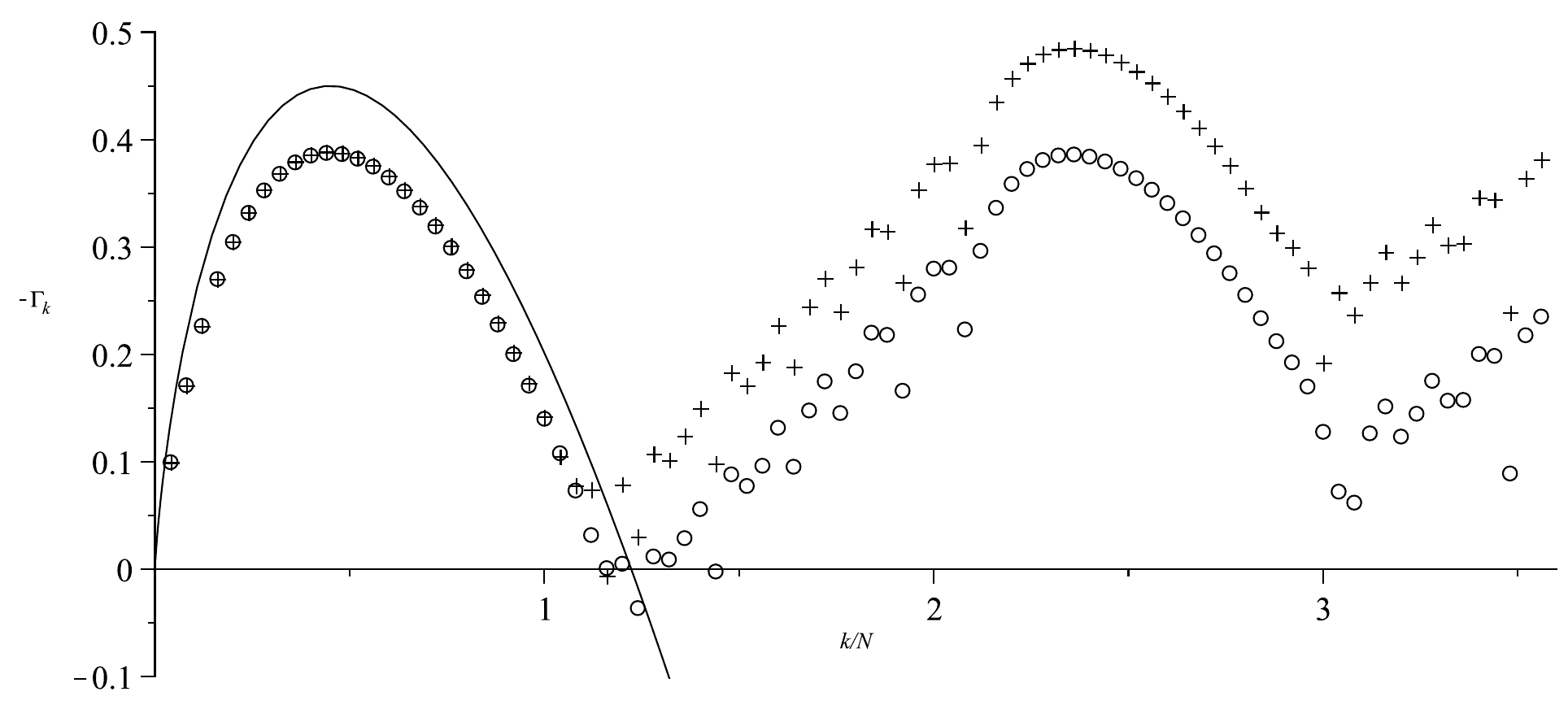}
\else
\includegraphics[scale=0.8]{fA.eps}
\fi
\caption{$-\Gamma_k=\ln \langle \mathrm{Tr}_{{\cal S}_k} U \rangle / N$ is plotted on the vertical
axis as a function of $k/N$ which is plotted on the horizontal axis,
in the strongly coupled Gross-Witten model with $p=0.45$ and $N=45$.
Crosses represent a computation where $U$ is treated as a random variable,
open circles represent a computation assuming the saddle point approximation and
the probe approximation, {\it i.e.} treating $U$ as a classical variable, but retaining discreteness
of the eigenvalue distribution. The solid black
line represents the result due to the continuum approximation,
equation (\ref{saddle-point-strong-coupling}).}
\label{fA}
\end{figure}

To test the adequacy of this expression, we have performed several numerical computations.
In one, we treat $U$ as a random variable, approximating the integrals in equation
(\ref{loop2}) by Gaussian integrals near the saddle point (the precise methodology
is described in Section \ref{sec:gaussian}).  In another, we rely on both the saddle
point approximation and the probe approximation (but not the continuum approximation),
treating $U$ as a strictly classical variable but maintaining discreteness of the eigenvalue distribution.  
The results, at $p=0.45$, are given in
Figure \ref{fA}.  Three things are evident in this Figure:
\begin{itemize}
\item There is a good match between the results of the saddle point
computation plus continuum distribution (\ref{saddle-point-strong-coupling})
and the numerics
for smaller values of $\frac{k}{N}$. This match ends abruptly at $\frac{k}{N}\sim 1.25$.
\item For larger values of $\frac{k}{N}$, the numerical computations agree with each other rather well, but
disagree with (\ref{saddle-point-strong-coupling}).  Agreement of the numerical computations suggests that the
existence of a saddle point and use of the classical matrix $\hat U$ are still valid, apparently for the
whole range of $\frac{k}{N}$ that we consider.  However, the continuum approximation which led to the result (\ref{saddle-point-strong-coupling}) must fail for values at and above
${k \over N}\sim 1.25$.
\item  Treating $U$ as a random variable and treating it as a classical variable
with the operator $\mathrm{Tr}_{{\cal S}_k} U$ in the probe approximation results in
differences of sub-leading order, consistent with equation (\ref{probe0}).
\end{itemize}
Thus the saddle point approximation and the probe approximation appear to be valid
for the computation of $\langle \mathrm{Tr}_{{\cal S}_k} U \rangle $,
but the continuum approximation does not. Nonetheless, validity of the saddle point and
probe approximations allow us to write, using equation (\ref{frobenius}),
\be
\langle \mathrm{Tr}_{{\cal S}_k} U \rangle  = \left \langle{1\over k!}\sum_{\sigma \in S_k} \left ( \prod_{j=1}^p \mathrm{Tr} 
 U^{l_j} \right ) \right \rangle~ 
\approx~
{1\over k!}\sum_{\sigma \in S_k} \left ( \prod_{j=1}^p \mathrm{Tr}~\hat U^{l_j} \right )~.
\label{S-frobenius}
\ee
In the last equality, it is legitimate to substitute the classical discrete eigenvalues.

We now draw the reader's attention to the strange feature of
$\langle \mathrm{Tr}_{{\cal S}_k} U \rangle$ as a function of $k$,
namely the recurrence, or approximate periodicity, which is seen in Figure \ref{fA}.
The remainder of this Introduction will discuss the origin and interpretation of
this feature, assuming the validity of the formula (\ref{S-frobenius}) above.

\begin{figure}
\begin{center}
\ifpdf
\includegraphics[scale=0.75]{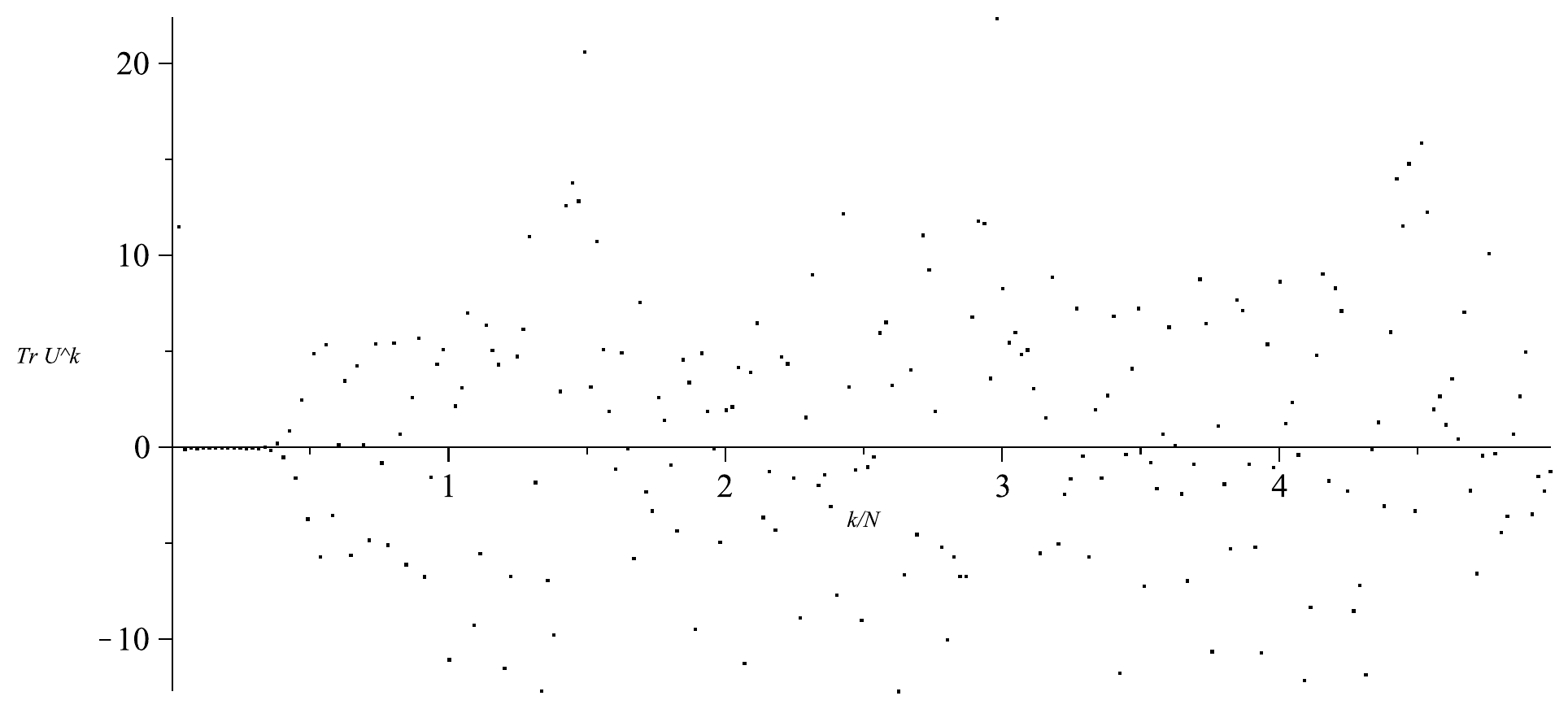}
\else
\includegraphics[scale=0.75]{f00a.eps}
\fi

(a)

\ifpdf
\includegraphics[scale=0.75]{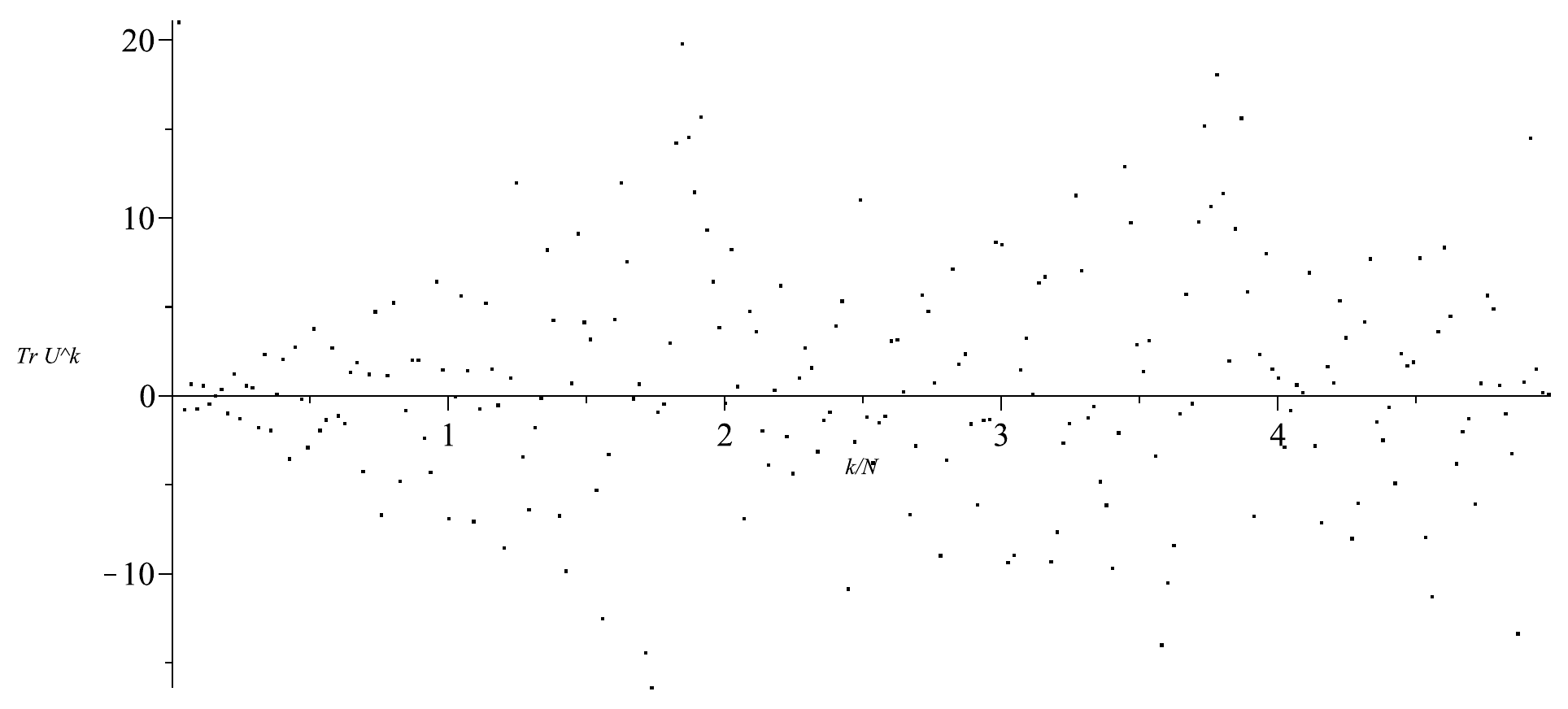}
\else
\includegraphics[scale=0.75]{f00b.eps}
\fi

(b)
\end{center}
\caption{Numerical simulation of the Gross-Witten model in the strong coupling
phase with $N=45$ and (a) $p=0.25$ and (b) $p=0.45$.  $\mathrm{Tr}~\hat U^k $ is plotted on the vertical axis
as a function of $k/N$ on the horizontal axis.}
\label{f00}
\end{figure}

Formula (\ref{S-frobenius}) implies that \emph{for the purpose of computing}
$\langle \mathrm{Tr}_{{\cal S}_k} U \rangle$, we can treat $U$ as a classical variable,
frozen at the saddle point value $\hat U$.  To study the implications of this
formula, we need to examine the properties of $\mathrm{Tr}~\hat U^{k}$ for
large $k$ and large $N$, in the regime where $k \sim N$.  We illustrate these
properties in Figure \ref{f00}.

For $p=0.25$, Figure \ref{f00}(a), the results shown agree with the continuum approximation result (\ref{grosswittentraces})
up to $k/N \approx 0.35$.  Above this point, $\mathrm{Tr}~\hat U^k $ is no longer
zero, as was predicted by equation (\ref{grosswittentraces}), and is instead of order $N$
(we should recall here that since $\hat U$ is a unitary matrix, $|\mathrm{Tr}~\hat U^k | \leq N$,
so above $k/N \approx 0.35$, $\mathrm{Tr}~\hat U^k $ is quite large, reaching as much as half of its
maximum value.).

A reasoning for why this is so comes from the fact that,
at any finite $N$, trace relations allow us to compute the traces of a single classical matrix
$\mathrm {Tr}~\hat U^k$ for $k>N/2$ from the first $N/2$ traces.
It is therefore not possible for just the first trace
to be arbitrary and non-zero, and the rest zero, as the continuum approximation would demand.

Similarly, for general (but assumed smooth) eigenvalue distributions with a infinite
number of frequency components, $\mathrm {Tr}~\hat U^k$
would have to decay to zero with $k\rightarrow \infty$, which again does
not appear compatible with trace relations. Thus, we might expect $\mathrm {Tr}~\hat U^k$
to become noisy at some point in the vicinity of $\frac{k}{N}\sim\frac{1}{2}$, and this
is indeed what we see for the case of $p=0.25$ in Figure \ref{f00}(a).  Notice, in Figure \ref{f00}(b), 
that at $p=0.45$ which is close to the maximum value that $p$ can reach in the strong coupling phase
of the Gross-Witten model (before the
phase transition to the weak coupling phase with a gapped distribution), the agreement with
the continuum approximation is even weaker.

More intuitively, consider the eigenvalues $\{\hat\phi_1,\ldots,\hat\phi_N\}$ and
the spacing between consecutive eigenvalues $\hat \phi_{i+1}-\hat\phi_i$.  At large $N$, this
spacing is approximately $( N \rho(\hat \phi_i))^{-1}$ where $\rho(\phi)$
is the eigenvalue density.  When summing over the phases $e^{ik\hat\phi_j}$, we are probing
the structure of the distribution of eigenvalues at wave-number $k$, and 
we cannot replace the discrete sum with an integral unless
$k\cdot($spacing$) \ll 1$, or $k/N \ll \rho$.  The continuum approximation
to the density will thus hold for a finite $k/N$, as long as it is not too large, the
cutoff being of order the inverse height of $\rho(\phi)$.
The narrower the  eigenvalue distribution, the larger the density
$\rho(\phi)$ and therefore the continuum approximation will hold
for a larger range of $k/N$.  Only for the delta-function eigenvalue
distribution will the continuum approximation hold for any $k/N$
of order $N^0$.

\begin{figure}
\ifpdf
\includegraphics[scale=0.75]{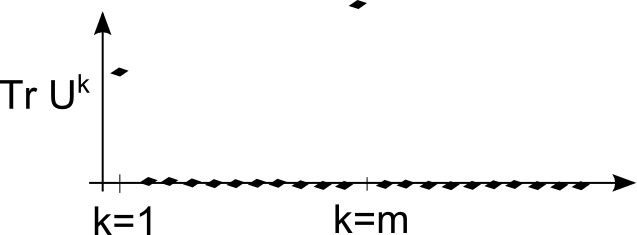}
\else
\includegraphics[scale=0.75]{f000.eps}
\fi
\caption{A caricature of Figure \ref{f00}.}
\label{f000}
\end{figure}

Returning to the strongly coupled Gross-Witten model, the continuum approximation
result (\ref{saddle-point-strong-coupling}), which is based on equation (\ref{grosswittentraces}),
implies that $\langle \mathrm{Tr}_{{\cal S}_k} U \rangle$ becomes very small once
$k/N$ is large enough.  This is a generic feature of $\langle \mathrm{Tr}_{{\cal S}_k} U \rangle$
which appears when $\mathrm {Tr}~\hat U^k$ goes to zero fast enough as $k \rightarrow \infty$.
Define $\gamma_k$ by $\langle \mathrm{Tr}_{{\cal S}_k} U \rangle = \exp(-N \gamma_k)$ in such
a scenario.  We can think of $\gamma_k$ as $\Gamma_k$ as obtained in the continuum approximation
and pictured with a solid line in Figure \ref{fA}.
As we have just discussed, at any finite $N$, $\mathrm {Tr}~\hat U^k$ cannot in fact
go to zero for large $k$, due to restrictions placed on it by the trace relations.  We will examine
the consequences of this fact by assuming that $\mathrm {Tr}~\hat U^k$ fails to be zero at precisely
one large value of $k$, which we will denote with $m$, $\mathrm {Tr}~\hat U^m = s \sim N$.  We will assume
that $m\sim N$. 
This is motivated by Figure \ref{f00}, where
we see that $\mathrm {Tr}~\hat U^k$ takes exceptionally large values at isolated points, the smallest
of which is $m = k \approx 1.8 N$ (for $p=0.45$, Figure \ref{f00}(b)).  What we are doing is
replacing the situation in Figure \ref{f00} with a caricature shown in Figure \ref{f000}. (To be clear, 
$\gamma_k$ could be computed by assuming that the only non-zero trace is ${\rm Tr}~\hat U^{\pm 1}$ whereas 
$\Gamma_k$ is computed assuming
that the only nonzero traces are ${\rm Tr}~\hat U^{\pm 1}$, ${\rm Tr}~\hat U^{\pm m}$, though
the argument is somewhat more general than that, as it relies on the  large $k$ behaviour of $\gamma_k$,
and not the details of ${\rm Tr}~\hat U^{k}$ for small $k$.)

Equation (\ref{frobenius}) now implies that, for $n$ of order $N^0$
and $k<m$,
\be
e^{-N\Gamma_{k+nm}}=\mathrm{Tr}_{{\cal S}_{k+nm}} \hat U = {k! \over (k+nm)!} \left [
\sum_{a=0}^n (Ns)^a {(k+nm)! \over a! m^a (k+(n-a)m)!} e^{-N \gamma_{k+(n-a)m}}
\right ]~.
\ee
Assuming that $m$ is large enough that $e^{-N \gamma_{k+nm}} \ll
e^{-N \gamma_{k+(n-1)m}} \ll \ldots \ll  e^{-N \gamma_{k}}$, the
$a=n$ term dominates the sum and we have
\be
\mathrm{Tr}_{{\cal S}_{k+nm}} \hat U \approx {(Ns)^n \over n! m^n} e^{-N \gamma_{k}}
= {1\over n!} \left (s \over m/N\right)^n  \mathrm{Tr}_{{\cal S}_{k}} \hat U~,
\label{periodicity}
\ee
or
\be
\Gamma_{k+nm} = \Gamma_k + {\cal O}(\ln(N))~.
\label{periodic}
\ee
We discover that the order $N$ part of $\Gamma_k$ is periodic in $k$, with the period
given by $m$.  This simple calculation
reproduces the behavior seen in Figure \ref{fA} quite well.  In particular,
we see that the period of the recurrence in Figure \ref{fA} is about $1.8$
which matches really well anomalously large value of ${\rm Tr}~\hat U^k$
seen at $k=1.8 N$ in Figure \ref{f00}(b). 
 
Going back to Figure \ref{fA} we see that $\mathrm {Tr}~\hat U^k$ is also
large at $k=2m$ and $k=3m$ (and probably at higher multiples).  How does that
affect our computation?  Let's say that $\mathrm {Tr}~\hat U^{2m} = \tilde s$.
Then, equation (\ref{periodicity}), for $n=2$, becomes
\be
\mathrm{Tr}_{{\cal S}_{k+2m}} \hat U \approx
\left ( {1\over 2!} \left (s \over m/N\right)^2
+{1\over 1!} \left (\tilde s \over m/N\right)^1 \right )
 \mathrm{Tr}_{{\cal S}_{k}} \hat U~,
\ee
which gives $\Gamma_{k+2m} = \Gamma_k + {\cal O}(1/N)$ as before.  Thus,
the presence of further points where $\mathrm {Tr}~\hat U^k$ is nonzero
does not affect our conclusions as long as these appear at $k$ equal
to integer multiples of $m$.  This turns out to be a generic feature of
simple eigenvalue distributions, and we will discuss it further in Section \ref{sec:gw}.
We will also examine a more complicated example with multiple periods in
Section \ref{sec:discussion}.

Before moving on, let us briefly discuss the origin of the values of $k$ for which 
$\mathrm {Tr}~\hat U^k$ is anomalously large.  They can occur when there is a region in the interval
$[-\pi,\pi]$ where a large number of the classical eigenvalues are almost equally spaced.
This would damp the destructive interference between the individual terms in the sum over $j$
of the phases $e^{ik\hat\phi_j}$.  It would occur for values of the wave-vector $k$ which are
$2\pi\cdot{\rm integer}/$(spacing).  Eigenvalues are almost equally spaced when the derivative of the
eigenvalue distribution is zero, that is, at an extremum of the distribution.  We will return to this
issue in the following Section and see a high degree of correlation between the period of recurrences and
the maxima of the eigenvalue distribution. 
This would also explain why the anomalously large ${\rm Tr}~\hat U^k$ 
seen in Figure \ref{f00} apparently occur for values of $k$ which are integer multiples of a basic number. 
 
\begin{figure}
\ifpdf
\includegraphics[scale=0.75]{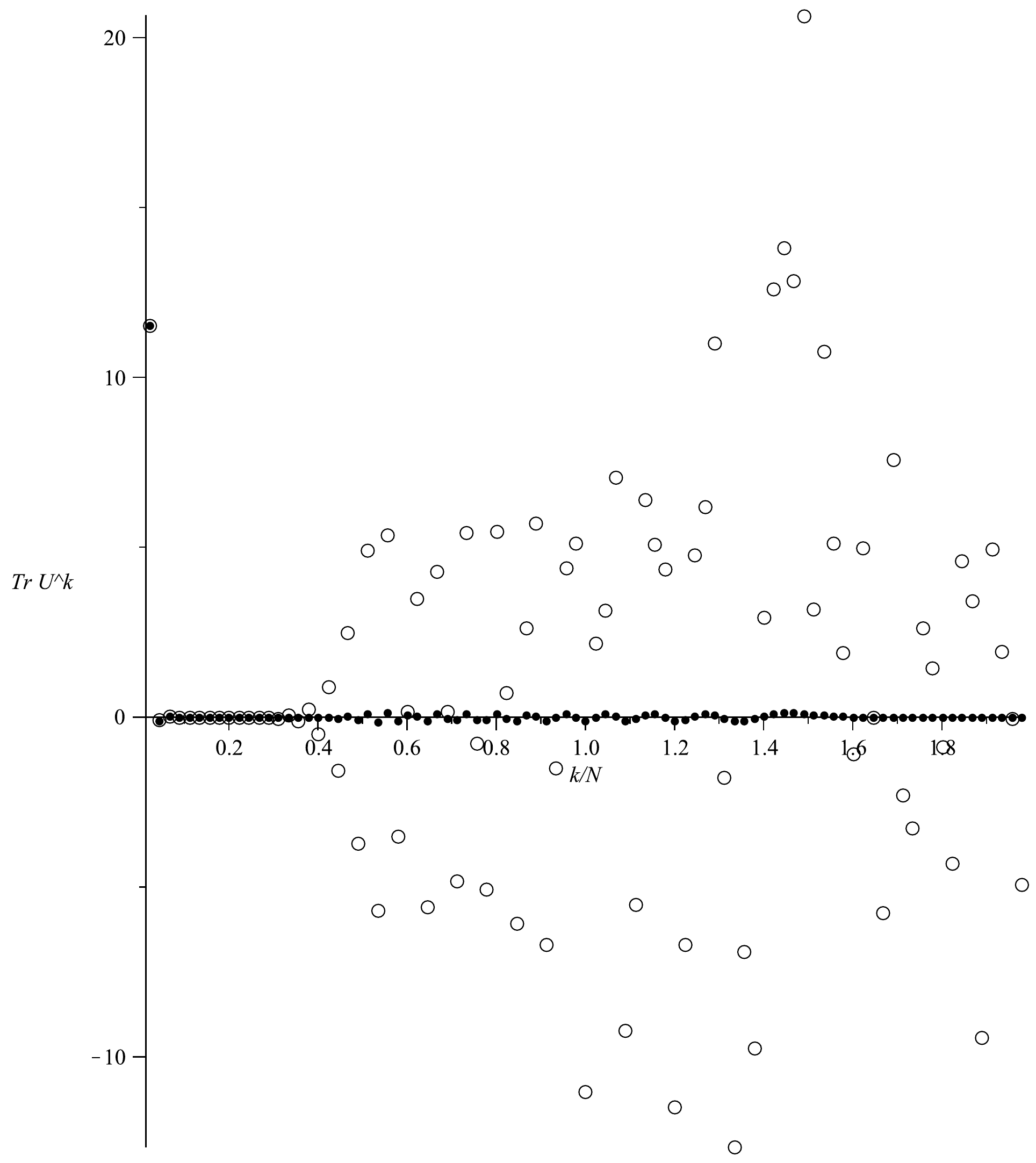}
\else
\includegraphics[scale=0.75]{f0.eps}
\fi
\caption{Numerical simulation of the Gross-Witten model in the strong coupling
phase with $p=0.25$ and  $N=45$.
$\langle \mathrm {Tr}~ U^k \rangle$ is plotted on the vertical axis
as a function of $k/N$ on the horizontal axis.
Open circles are obtained by finding $\hat\phi_i$ that minimize the effective action
and evaluating $\mathrm {Tr}~ \hat U^k$ using those eigenvalues.  This is
identical to Figure \ref{f00}(a).  Filled circles include the leading correction to
the large $N$ result, arising from integrating the fluctuations about $\hat\phi_i$.
The precise methodology is given in Section \ref{sec:gaussian}.
}
\label{f0}
\end{figure}

So far we have focused on the properties of $\mathrm {Tr}~\hat U^k$.
What about $\langle \mathrm {Tr}~ U^k \rangle$?  Figure \ref{f0}
addresses this question.  It shows  $\langle \mathrm {Tr}~ U^k \rangle$  in the Gross-Witten model
with $p=0.25$ computed by treating $U$ as a random variable, correcting
the classical limit by integrating over quadratic fluctuations around the saddle point (filled circles),
as well the classical limit itself, $\mathrm {Tr}~ \hat U^k$ (from Figure \ref{f00}(a)) for comparison.
Curiously, the filled circles do agree with the large $N$ result (\ref{grosswittentraces}) over the entire range.
We note that this integration is done for $SU(N)$, rather than $U(N)$ by imposing
the constraint that the average value
of the eigenvalues must be zero.  If we use the unconstrained integral over $U(N)$, there is a soft
mode which changes the result considerably.  We take the upshot of this discussion as evidence that the
saddle point approximation cannot be trusted for the computation of
$ \langle\mathrm {Tr}~ U^k\rangle$ when $k$ is bigger than
some fraction of $N$.  The fluctuations are apparently at least as important as
the classical eigenvalues in that regime.  This is consistent with equations
(\ref{saddle-point})  and (\ref{probe0}), since $ \langle\mathrm {Tr}~ U^k\rangle$ is
bounded by $N$ and therefore its logarithm cannot be large.
The failure of the saddle point approximation
arises from the variance in the
eigenvalues $\phi_i$ being of the same order as the spacing between $\hat \phi_i$s in the
classical limit.  Thus integrating over the eigenvalues `washes' out the effect of
a discrete spectrum and seems to restore the validity  of the continuum approximation for
$ \langle\mathrm {Tr}~ U^k\rangle$.
Another way to view this is that re-introducing some randomness to the matrix so that it is no
longer strictly classical relaxes the trace relations.  In fact, we see that this works rather
well in this case, in that fluctuations allow the traces to agree with what one would obtain from
the eigenvalue density in the large $N$ limit over the entire range of $\frac{k}{N}$ that we explore.

The curious (and central to our argument) fact is that this is \emph{not}
the case with $\langle \mathrm{Tr}_{{\cal S}_k} U \rangle$.
If we were to use $\langle \mathrm{Tr}_{{\cal S}_k} U \rangle$ instead of
$\mathrm {Tr}~ \hat U^k$ in equation (\ref{S-frobenius}), we would not
obtain the correct answer.  This can potentially be explained by a failure
of factorisability (it is unlikely that
$\prod_{k_i} \langle \mathrm{Tr}~U^{k_i} \rangle =
\left  \langle \prod_{k_i} \mathrm{Tr}~ U^{k_i} \right \rangle$
if $ \langle\mathrm {Tr}~ U^k\rangle$ cannot be accurately computed in the
saddle point approximation), but we do not explore this any further here.
In contrast to $ \langle\mathrm {Tr}~ U^k\rangle$, $\langle \mathrm{Tr}_{{\cal S}_k} U \rangle$
seems to be computable in the saddle point approximation
(as long as we do not use the continuum approximation as well) and to have
a well defined large $N$ limit, which we will explore further in
Section \ref{sec:gw}.


The periodic behavior of equation (\ref{periodic}) has an
interesting implication for some physical applications of this matrix model.  
For example, when the character is the expectation value of the Polyakov loop in 
Yang-Mills theory on the sphere,  $\Gamma_k$ is interpreted as the free energy
of a heavy quark with center charge $k$ and in the totally symmetric representation. 
If one thinks of this heavy quark as being composed of $k$ partons, it appears that $m$ of these partons can combine together
to form a (`bound') state of low free energy.  Thus, if $k=nm+\tilde k$
($0\leq \tilde k < m$), the free energy of $k$ partons receives contributions mainly from
the $\tilde k$ partons that are `free', while the contribution of the $n$ `bound' states of
$m$ partons is negligible.  We will argue that this periodicity always exists, since it is related
to maxima of the eigenvalue distribution. 

The rest of the paper is organized as follows.  In
Section \ref{sec:gw}, we study further the character
and its large $N$ limit in the Gross-Witten model,
both in strong and in weak coupling. We find that the character
is indeed quasi-periodic and link the period to the
maximum value taken by the eigenvalue density.
In Section \ref{sec:gaussian}, we numerically compute the integrals
in equation (\ref{loop2}) to test approximations 1 and 2,
and to see whether our results from Section \ref{sec:gw} are robust.
Finally, in Section \ref{sec:discussion}, we compute
$\Gamma_k$ for more complicated
eigenvalue densities and discuss generic behaviour.

\section{Character in the Gross-Witten model}
\label{sec:gw}

\begin{figure}
\ifpdf
\includegraphics[scale=0.75]{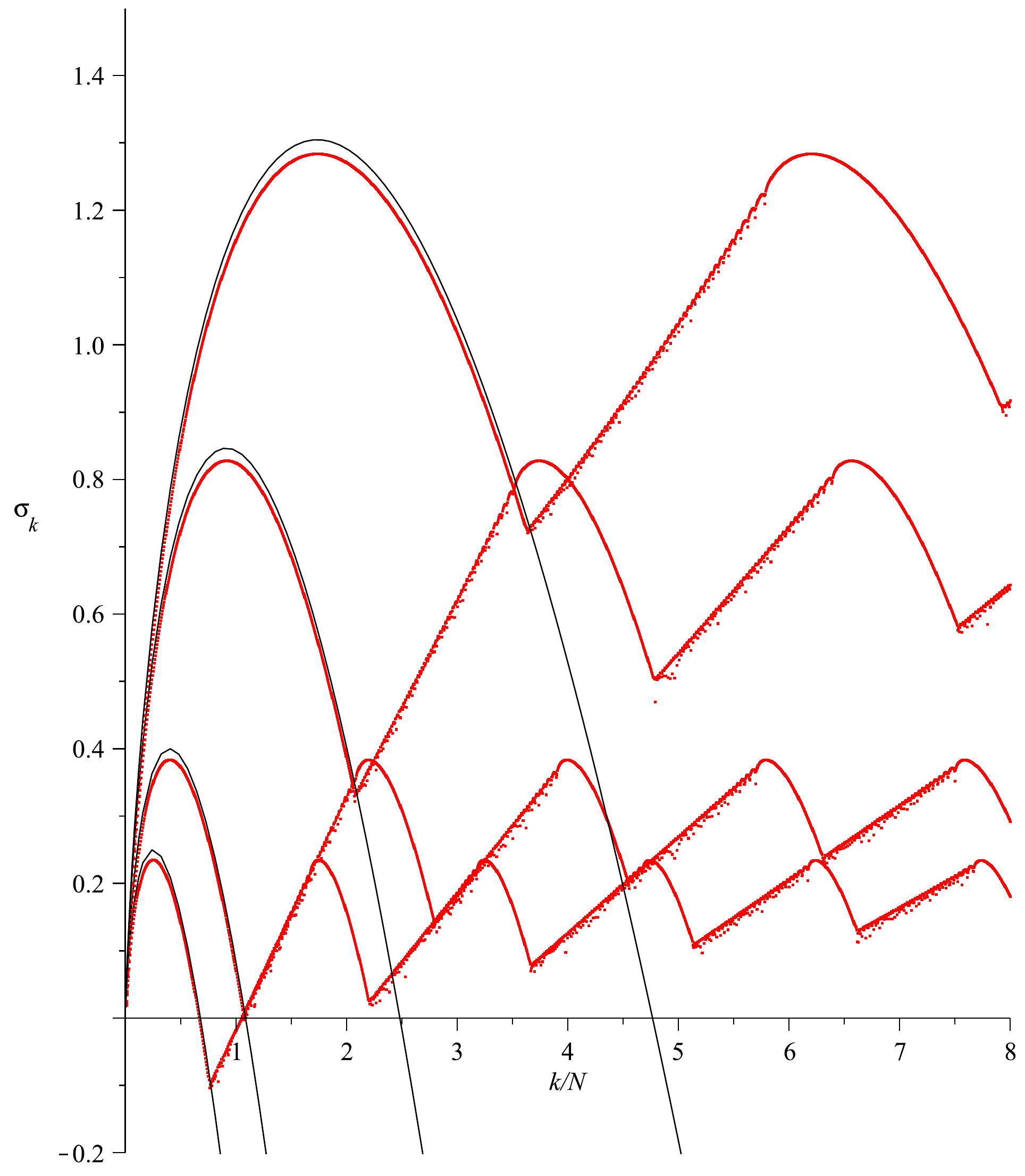}
\else
\includegraphics[scale=0.75]{f1.eps}
\fi
\caption{Numerical simulation of the Gross-Witten model in
both the strong and the weak coupling phases.
$\sigma_k$ on the vertical axis is plotted as a function of $k/N$ on the horizontal axis
for p=0.25, 0.40, 0.75, 0.90 (from the shortest recurrence period to the longest).
Red dots represent numerical results at N=200 and the solid
black line is the result in the continuum approximation.}
\label{f1}
\end{figure}

In this Section, we shall examine the difference between using a discrete and
continuous eigenvalue distribution in the limit where $N$ is large. We will assume
that the large $N$ limit localizes the matrix integral onto eigenvalues $\hat\phi_i$
and that to a first approximation, these eigenvalues are not disturbed by the presence
of the character in the integral.   To aid numerical computation at large $N$,
will use an approximation to the eigenvalues $\hat\phi_i$
which we obtain be re-discretizing the continuum distribution.  We will show that this is in fact a
good approximation in the next Section.  Here, it will prove sufficient
to demonstrate our point about the consequences of the failure of the
continuum approximation. To
be concrete, we will study the expectation value of a character in the
Gross-Witten model \cite{Gross:1980he}, both below and above the third-order
phase transition.

At finite $N$, we approximate
the actual positions of the eigenvalues, $\hat \phi_i$, by
approximate positions $\tilde \phi_i$  computed from
the large-$N$ density, via the formula which (at infinite $N$) defines the density,
\be
\label{eigenvalues}
\int_{-\pi}^{\tilde \phi_k} \rho(\phi) ={k-{1\over 2}\over N}~~~~
\mathrm{for~} k=1\ldots N~.
\ee
If $\rho(\phi) = \rho(-\phi)$, the ${1\over 2}$ simply leads to
a symmetric distribution of eigenvalues.  $\tilde \phi_i$s differ
from  $\hat \phi_i$s by corrections which are sub-leading in the
large-$N$ limit. We will demonstrate that they are a good enough
approximation to $\hat\phi_i$ in the next Section.

We will find it convenient to use the following generating
function for the character:
\be
\label{genfun}
\sum_{k=0}^\infty t^k ~\mathrm{Tr}_{{\cal S}_k} U  =
\prod_{j=1}^N {1\over 1-t e^{i \phi_j}} =
\exp \left (-\mathrm{Tr} \ln (1-tU)\right )~.
\ee
Then, $\sigma_k$ defined by
\be
\label{sigma}
\sum_{k=0}^\infty t^k e^{N\sigma_k}  =
\exp \left (-N \sum_{j=1}^N \ln (1-t e^{-\tilde \phi_j})\right )
\ee
is approximately equal to $-\Gamma_k$.
We will see in the next Section that the difference between
$\sigma_k$ and $-\Gamma_k$ is of order $1/N$.

In the ungapped phase of the Gross-Witten model, the
eigenvalue density is
\be
\rho(\phi) = {1\over 2\pi} (1+2p\cos \phi)
\ee
for $0<p<1/2$, while in the gapped phase, the eigenvalue
density is
\be
\rho(\phi) = {\cos{\phi \over 2} \over \pi (2-2p)} \sqrt{
2-2p-\sin^2(\phi/2)}
\ee
for $1/2<p<1$. Here, in both cases, we have parameterized the distributions
by the expectation value $p=\frac{1}{N}\langle{\rm Tr}~U\rangle$.
With these explicit eigenvalue densities, at finite fixed $N$,
we solve equation
(\ref{eigenvalues})  (numerically) for $\tilde \phi_i$ at different values of $p$.
This allows us to write down the generating function (\ref{sigma}) as
a function of $t$, which we then expand (using Maple) in a Taylor series
for small $t$ to obtain $\sigma_k$
as a function of $k$.  The results, at N=200, are displayed in Figure
\ref{f1}.  Since both axes of the Figure are scaled so that
the plots are independent of $N$ (to leading order),  finite
$N$ results should approximate  infinite $N$ results.
For comparison, solid lines show the answer at infinite $N$
in the continuum approximation, as computed in the introduction and in
\cite{Grignani:2009ua}.
The approximate periodicity of $\sigma_k$ as a function of $k/N$,
predicted in the Introduction, is clearly visible in the Figure.

\begin{figure}
\ifpdf
\includegraphics[scale=0.75]{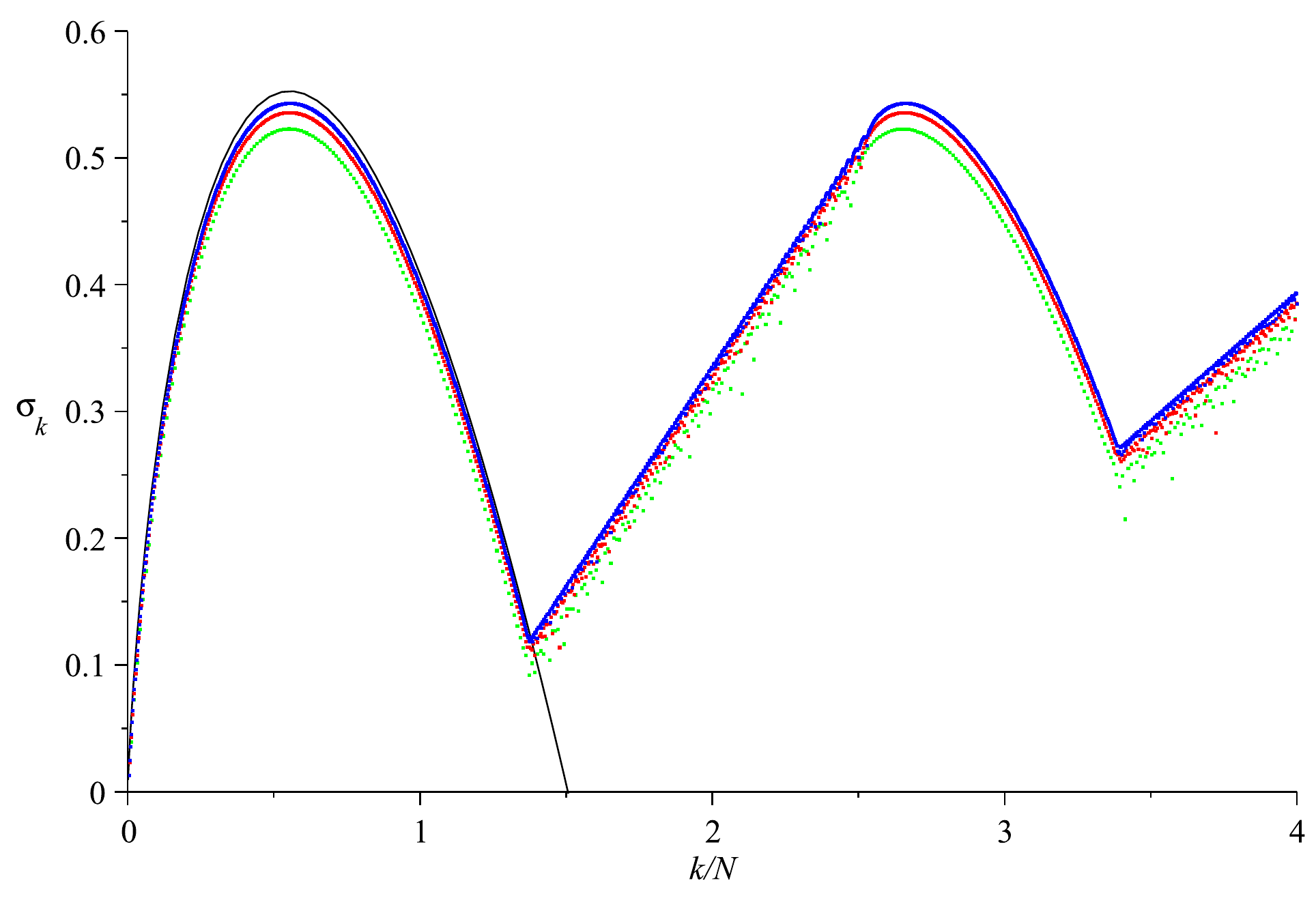}
\else
\includegraphics[scale=0.75]{f2.eps}
\fi
\caption{Numerical simulation of the Gross-Witten model for $p=0.55$.
$\sigma_k$ on the vertical axis is plotted as a function of $k/N$ on the horizontal axis
for N=100 (green), N=200 (red) and N=400 (blue).  The solid
black line is the result in the continuum approximation.}
\label{f2}
\end{figure}

One might worry that the recurrences shown in Figure \ref{f1} are
simply a consequence of working at finite $N$ and do not accurately
represent the large $N$ limit of the theory.  To show that this is
not the case, Figure \ref{f2} contains the results of a computation
at $N=100$, $200$ and $400$.  Finite $N$ results seem to converge
to a well-defined answer with a stable recurrence period.

\begin{figure}
\ifpdf
\includegraphics[scale=0.85]{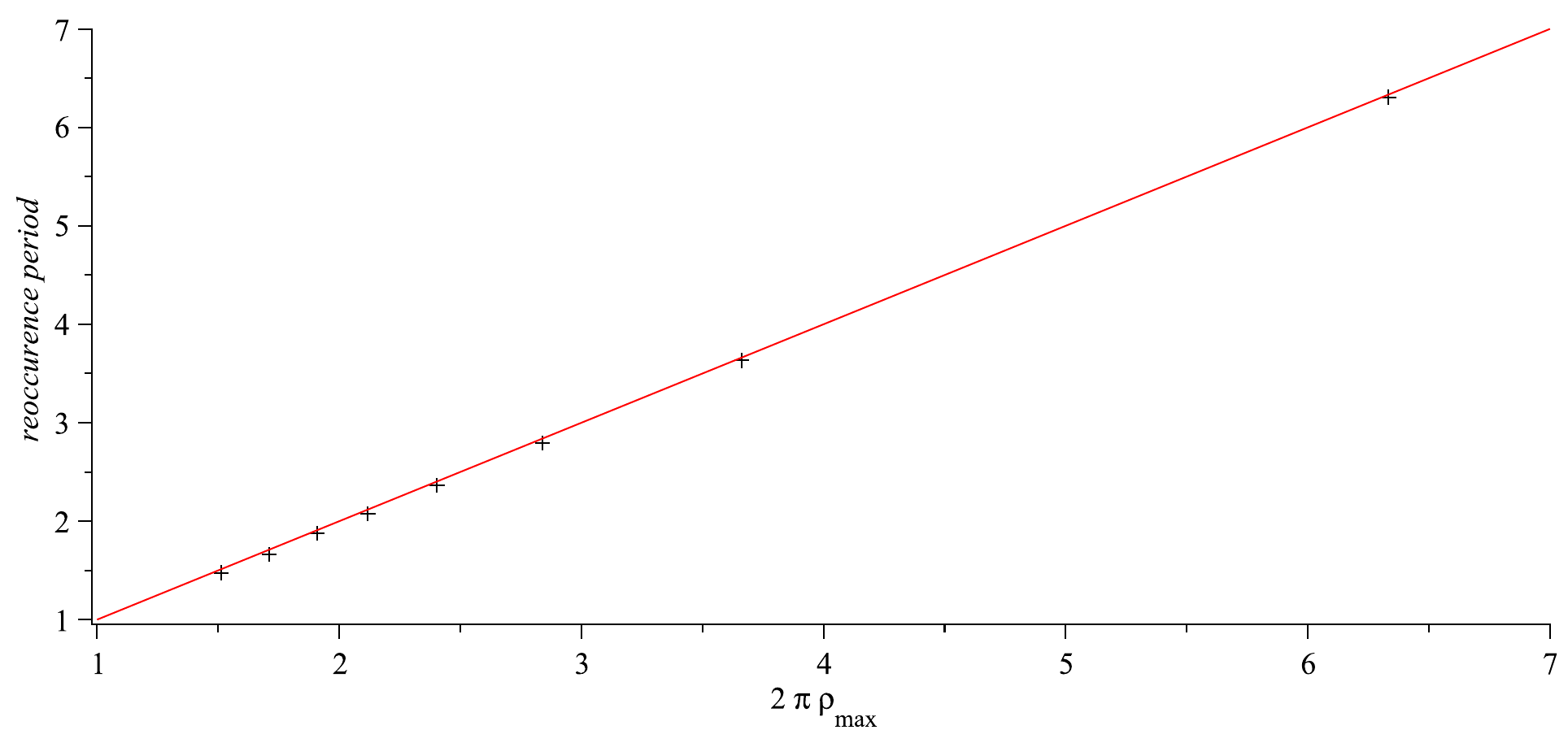}
\else
\includegraphics[scale=0.85]{f3.eps}
\fi
\caption{The recurrence period as a function of $2 \pi \rho_{max}$.
The crosses are data based on our numerical results and the solid line
is our prediction that recurrence period = $2 \pi \rho_{max}$.}
\label{f3}
\end{figure}

We can understand the period of the recurrences as follows.  In the
Introduction, we argued that the recurrences are due to a single
large trace, $\mathrm{Tr}~\hat U^m$ at some $m$.  Since
$\mathrm{Tr}~U^m = \sum_j \exp(im\tilde\phi_j)$, $\mathrm{Tr}~\hat U^m$
is large if the spacing between consecutive $\phi_j$ is an integer
multiple of $2\pi/m$.  Of course, the spacing is not constant, so what
we want is that a large fraction of the eigenvalues be spaced at
approximately $2\pi/m$.  The eigenvalue spacing,
equal to $(N \rho(\phi))^{-1}$ varies slowest where
the derivative of $\rho(\phi)$ is zero.  Since there are more eigenvalues
near the point where $\rho(\phi)$ attains its maximum than near the point
where it attains its minimum, we conjecture that $m$ is given by
$2\pi N \rho_{max}$, where $\rho_{max}$ is given by
\be
\rho_{max} = \left \{ \begin{array}{ll}
{1+2p \over 2\pi} & 0<p<1/2~, \\
{\sqrt{2-2p}\over 2\pi(1-p)} & 1/2 <p<1~.
\end{array}\right .
\ee
The recurrence period for $\sigma_k$ as a function of $k/N$ should then be
simply $2\pi\rho_{max}$.  This conjecture is clearly
supported by Figure \ref{f3}.
We will see in Section \ref{sec:discussion} that for more complicated
eigenvalue densities with multiple maxima, there will be several
competing recurrence periods.  Nonetheless,  the periods will be related
to the local maxima of the eigenvalue density in the way described above.

\begin{figure}
\ifpdf
\includegraphics[scale=0.85]{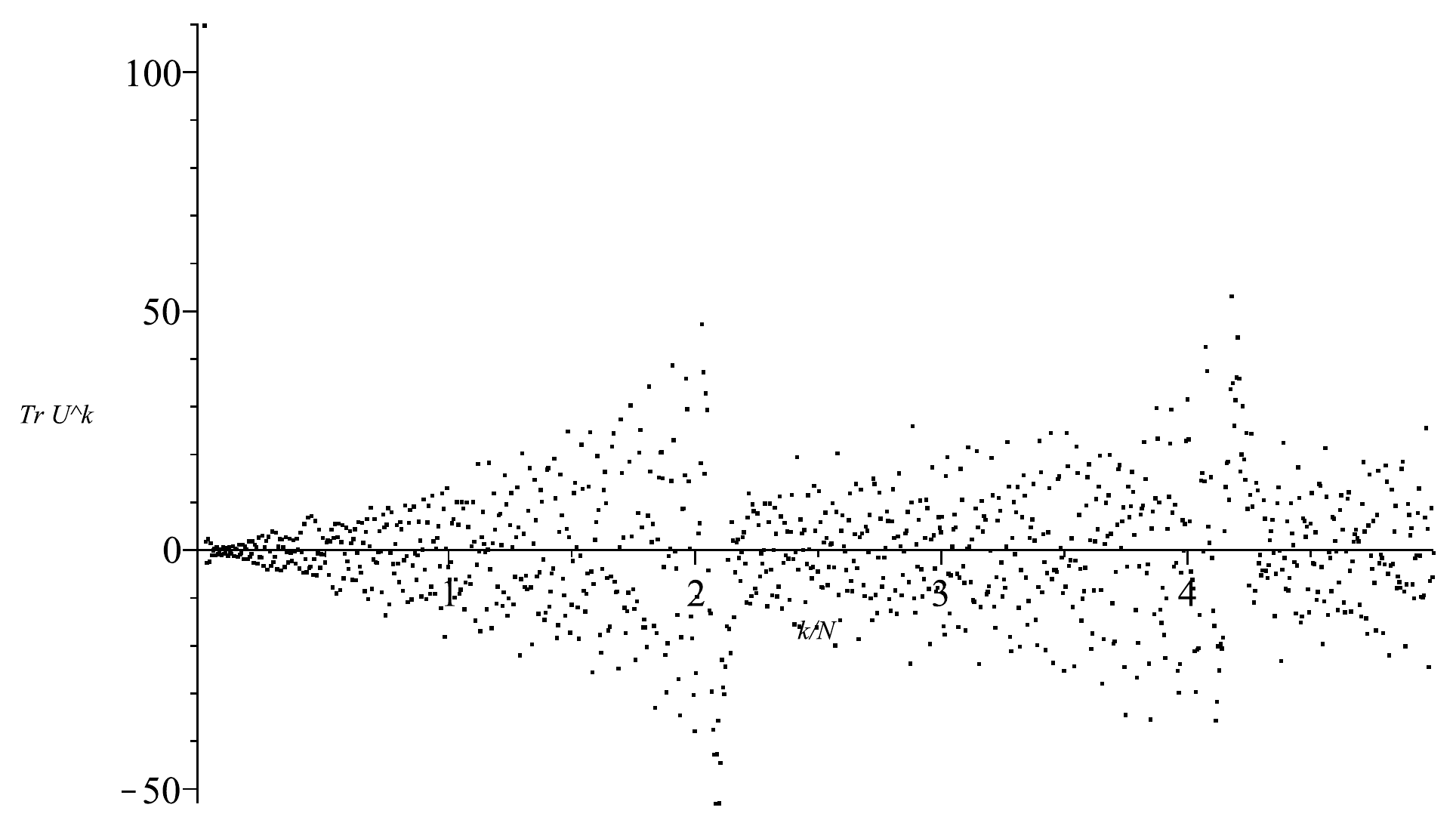}
\else
\includegraphics[scale=0.85]{f4.eps}
\fi
\caption{$\mathrm{Tr} ~\tilde U^k$ in the Gross-Witten model for $p=0.55$ and $N=200$.}
\label{f4}
\end{figure}

Of course, the toy picture painted in the Introduction, where there
$\mathrm{Tr} ~\hat U^m$ is large for only one isolated value of $m$ is not
accurate even in the case of the relatively simple eigenvalue distributions
discussed in this Section.  This was already illustrated in Figure \ref{f00}.
Here we complement that Figure with a computation in the weak coupling regime.
Figure \ref{f4} shows $\mathrm{Tr} ~\hat U^k$
for $p=0.55$ and $N=200$.  Several things are apparent in this plot:
$\mathrm{Tr} ~\hat U$ is large, while $\mathrm{Tr} ~\hat U^k$  is small
at first, but grows with $k$.  It also appears nearly random with a distribution
whose mean is zero.  Since $\mathrm{Tr} ~\hat U^k$ oscillates, the effect
described in the Introduction from each individual $\mathrm{Tr} ~\hat U^k$
roughly cancels between different values of $k$.
However, at $k/N\approx 2.1$ something interesting happens:
$\mathrm{Tr} ~\hat U^k$ is of order $N$ and negative for several values of $k$
in a row.  Thus, the effect for several values of $k$ in a row can reinforce, which
should lead to a recurrence.  We see in Figure \ref{f2} that indeed, at $p=0.55$,
the recurrence period is about 2.  Not surprisingly, a similar reinforcement
will occur for $k/N \approx 4.2$.

\section{Matrix model beyond the saddle point approximation}
\label{sec:gaussian}

In this Section we provide the methodology behind Figures
\ref{fA} and \ref{f0} in the Introduction and
further support the validity of our approximation in
Section \ref{sec:gw}.
While we have shown already that the recurrence pattern
is stable in the large $N$ limit, one could wonder whether it
is also stable under small (of order $1/N$) corrections to the
eigenvalue density.  We have checked the impact of such order $1/N$
changes as removing a single eigenvalue and adding a small high
frequency component to the eigenvalue density, and these had no
impact on the overall pattern beyond $1/N$ corrections.

Another possibility is that the since the eigenvalues are
random variables with a variance of order $1/N$, the discreteness
of the eigenvalue distribution might be washed out when the
integrals in equation (\ref{loop2}) are evaluated.
To test for this possibility, in this Section we evaluate these integrals
by approximating the potential near
the saddle point (to the leading, quadratic order).
While this computation is not exact, it should provide us with
the next-to-leading order corrections to $\sigma_k$ and
allow us to see whether the discreteness of the eigenvalue
density still matters when the eigenvalues are random variables.

The expectation value we are interested in can be written as
\be
\langle e^{-\mathrm{Tr} \ln(1-tU)} \rangle =
{\int \prod_{j=1}^N dx_j~
e^{-V_{\mathrm{eff}}(x_j)} ~
\exp \left (-\sum_j \ln(1-te^{ix_j}) \right )
\over
\int \prod_{j=1}^N dx_j~
e^{-V_{\mathrm{eff}}(x_j)}}~,
\label{veff}
\ee
where the effective potential for the eigenvalues is
\be
V_{\mathrm{eff}} = -\sum_{i<j} \ln \left [
\sin^2 \left ( {x_i-x_j}\over 2\right ) \right ]
~-~ 2 p N \sum_i \cos(x_i)~.
\ee
The first term in the potential, which comes from the Vandermonde
determinant, causes eigenvalues to repel, while
the second term attracts them towards $x_j=0$.
The second term comes from the Gross-Witten action (\ref{grosswitten})
where $\lambda = 1/p$ for $p<1/2$ and $\lambda = 4-4p$ for $p>1/2$.
This action is known\cite{Gross:1980he} to produce the eigenvalue distributions discussed
in the Section \ref{sec:gw}.

We will denote by
$\hat x_i$ the positions of the eigenvalues for which $V_{\mathrm{eff}}$
attains its minimum.  $\hat x_i$ are, up to corrections of order $1/N$, the
same as $\tilde \phi_i$ in equation (\ref{eigenvalues}).
The minimum is unique up to a permutation of the eigenvalues $x_i$,
and satisfies $\sum \hat x_i = 0$.
Given $\hat x_i$, we can expand $V_{\mathrm{eff}}$ around its minimum
\bear
V_{\mathrm{eff}}(x_1, \ldots, x_{N} ) &\approx&
V_{\mathrm{eff}}(\hat x_1, \ldots, \hat x_{N} ) \\ &+&
\half \sum_{a,b=1}^{N}
{\partial^2 \over \partial x_a \partial x_b}
V_{\mathrm{eff}}( x_1, \ldots, x_{N} )
|_{x_k = \hat x_k}~
\left (x_a - \hat x_a \right) \left (x_b - \hat x_b \right)~.
\eear

\begin{figure}
\ifpdf
\includegraphics[scale=0.75]{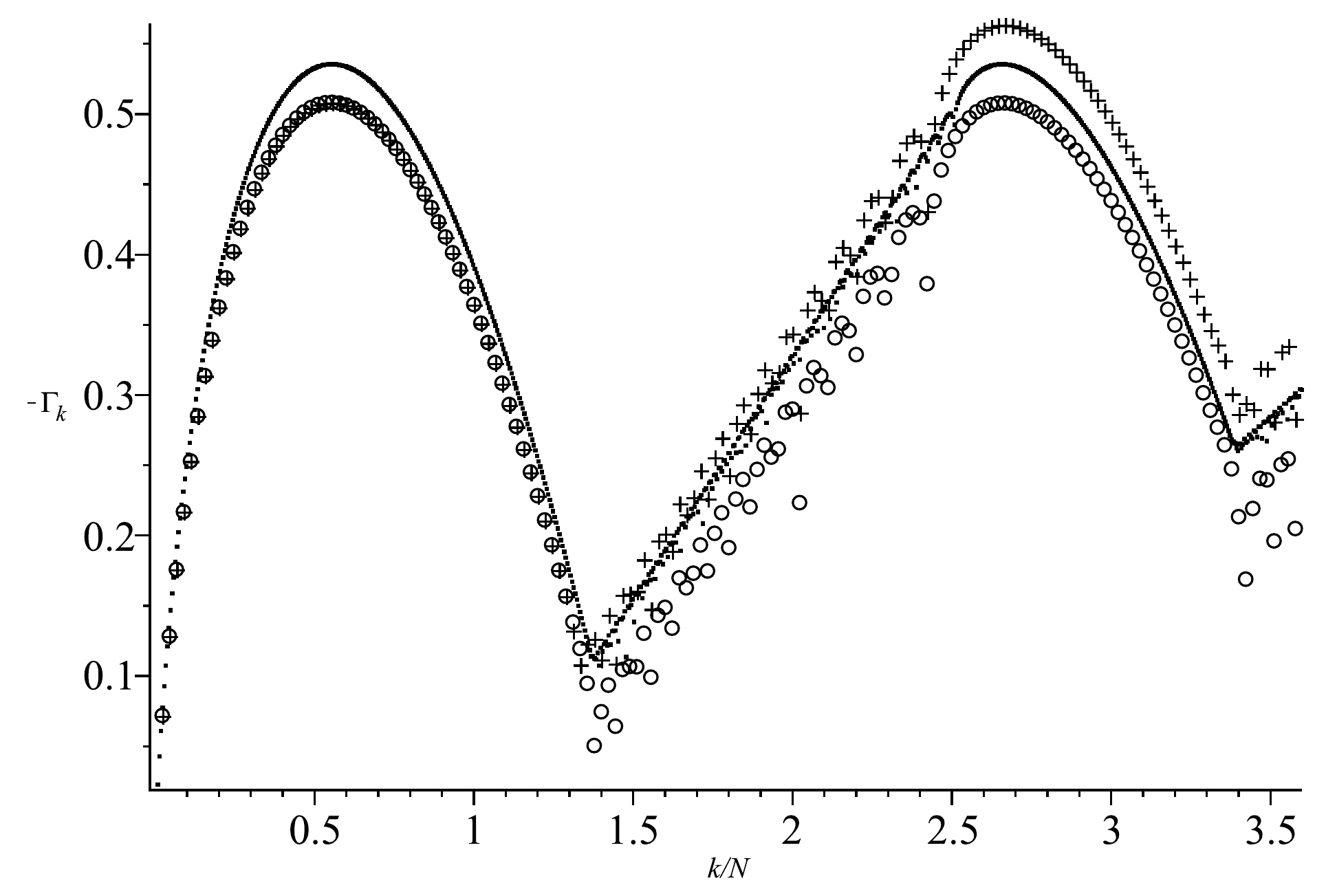}
\else
\includegraphics[scale=0.8]{f5.eps}
\fi
\caption{$-\Gamma_k$ in the Gross-Witten model as a function of $k/N$ for $p=0.55$.
Open circles represent the saddle point result
(the first term in equation (\ref{gaussian-result})), while crosses
include $1/N$ corrections (both terms in
equation (\ref{gaussian-result})); N=45.  The black points show the result of
an approximate saddle point computation with N=200 as discussed in
Section \ref{sec:gw}, for comparison.  The three plots differ by $1/N$
corrections.
}
\label{f5}
\end{figure}

We also expand  the inserted operator
$O = \exp \left (-\sum_j \ln(1-te^{ix_j}) \right )$
around the point $x_i=\hat x_i$,
\bear
O (x_1, \ldots, x_{N} ) &\approx& O (\hat x_1, \ldots, \hat x_{N} )
+ \sum_{a=1}^{N} {\partial \over \partial x_a} O (x_1, \ldots, x_{N} )
|_{x_k = \hat x_k}~ \left (x_a - \hat x_a \right)
\\ &+&
\half \sum_{a,b=1}^{N}
{\partial^2 \over \partial x_a \partial x_b}
O( x_1, \ldots, x_{N} ) |_{x_k = \hat x_k}~
\left (x_a - \hat x_a \right) \left (x_b - \hat x_b \right)~.
\eear
The second term on the first line does not contribute to the integrals,
by symmetry.  Evaluating the appropriate Gaussian integrals we obtain
\be
\langle e^{-\mathrm{Tr} \ln(1-tU)} \rangle =
O (\hat x_1, \ldots, \hat x_{N} ) ~+~
\half \mathrm{Tr} \left [
\left ( {\partial^2 O \over \partial x_a \partial x_b}\right )
\left ( {\partial^2 V_{\mathrm{eff}} \over \partial x_a \partial x_b}\right )^{-1}
\right ]~,
\label{gaussian-result}
\ee
where we have arranged the second-order derivatives into $N \times N$
matrices in the natural way.  $\hat x_i$, and the matrices of
second-order derivatives can be obtained in Maple.
The resulting expectation value is a
function of $t$, and can be expanded at small $t$ as before.  The first
term in the above equation is simply the saddle point result we
have obtained before (but with $\hat x$ instead of $\tilde \phi$),
while the second term represents $1/N$ corrections due to variance in the
eigenvalues.  As can be seen in Figure \ref{f5} (and in Figure \ref{fA}),
the general pattern we observed in
the previous Section persists.  The difference
between the saddle point results at $N=45$ (open circles) and at $N=200$
(points) is of order $1/N$, as is the difference between the result of
a Gaussian integral (crosses) and just the saddle point (open circles).
This again justifies approximations 1 and 2 described in the Introduction,
demonstrates that our methodology in Section \ref{sec:gw}
is sufficient and shows that the recurrences we have seen
appear in the exact computation as well.

To compute the results shown in Figure \ref{f0}, we proceed in a similar
fashion, except that we do not expand the operator, $\sum_j \exp(ik x_j)$
around $x_j = \hat x_j$, as it is not necessary to do that to obtain the
integrals.  Also, we impose a constraint $\sum x_i = 0$ in our integrals,
corresponding to the $SU(N)$ model, as the $U(N)$ model has a soft mode
which leads to $\langle \mathrm{Tr}~U^k\rangle \approx 0$ even at small $k$
(including $k=1$).

\section{Further examples and discussion}
\label{sec:discussion}

\begin{figure}
\ifpdf
\includegraphics[scale=0.3]{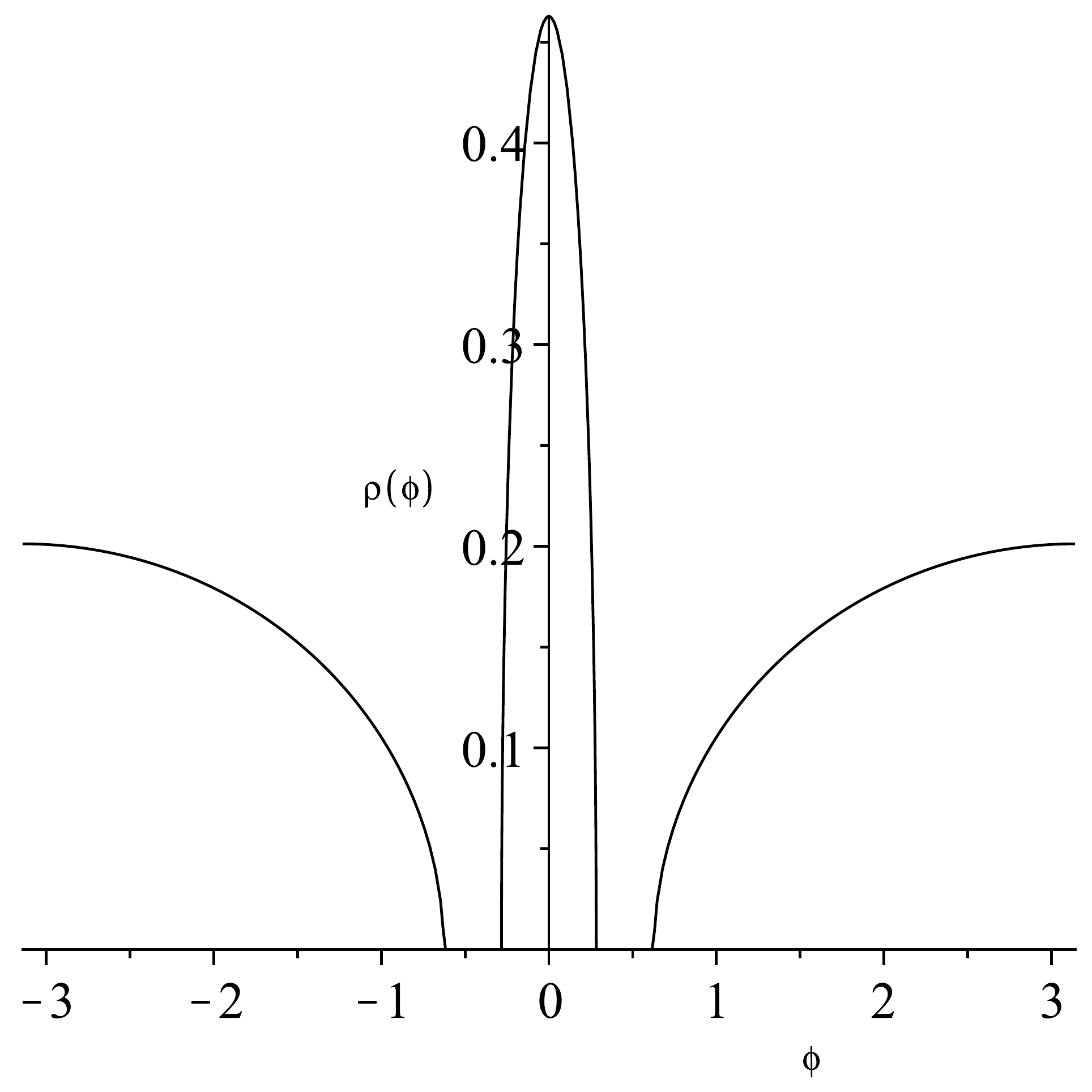}
\else
\includegraphics[scale=0.3]{f6-a.eps}
\fi
\ifpdf
\includegraphics[scale=0.5]{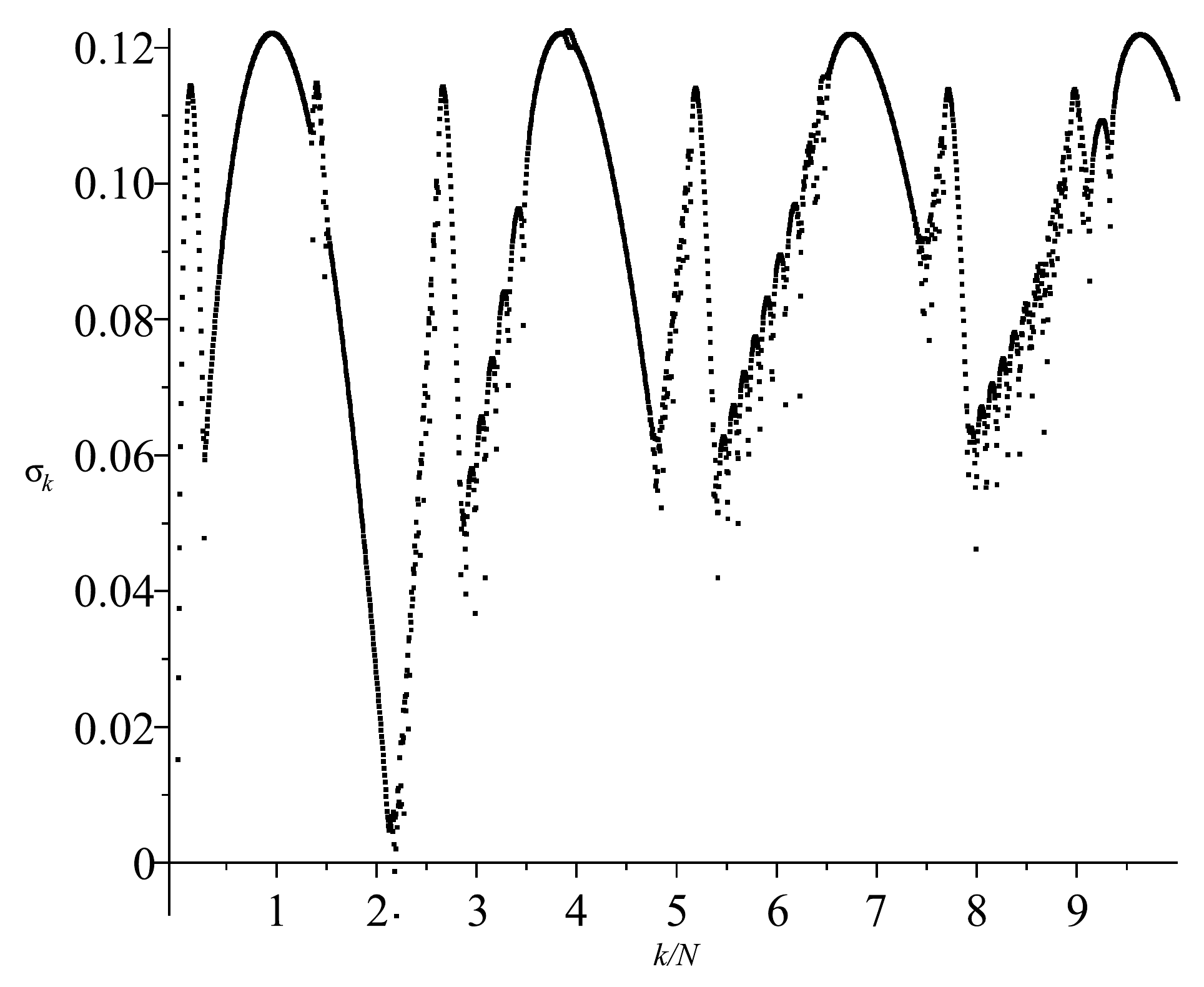}
\else
\includegraphics[scale=0.5]{f6-b.eps}
\fi
\caption{$\rho(\phi)$ given in equation (\ref{two-semi})
and the corresponding  $\sigma_k$ as a function of $k/N$,
for $N=200$.}
\label{f6}
\end{figure}

The results presented in Section \ref{sec:gw} are quite generic
if one focuses on eigenvalue distributions with a single maximum
(either ungapped or one-cut gapped distributions).  When multiple maxima are present, however,
the behaviour can be much richer, as multiple maxima can lead to
multiple recurrence periods.  As an example, consider the following
two-cut distribution, consisting of two rescaled semicircle
pieces, one centered around $\phi=0$ and one around $\phi=\pi$
\be
\rho(\phi) = \left \{ \begin{array}{ll}
h_1 \sqrt{\phi^2-w_1^2}  & \mathrm{for~} |\phi|< w_1~, \\
h_2 \sqrt{(\phi-\pi)^2-w_2^2}  & \mathrm{for~} |\phi-\pi|< w_2~, \\
\end{array} \right .
\label{two-semi}
\ee
where $h_1= 0.46283$, $h_2= 0.20123$, $w_1 = 0.09\pi$ and $w_2 = 0.8\pi$.  The eigenvalue
distribution together with the corresponding $\sigma_k$ as a function of $k/N$ are
shown in Figure \ref{f6}.  Two different recurrence periods
are clearly visible, each related to one of the two maxima
by the formula (recurrence period) = $2\pi\rho_{max}$.  The eigenvalue
distribution was carefully chosen to make the amplitude of the
two series of recurrences approximately equal so that they would both
be visible.  Generically, even if the distribution has multiple maxima,
one of the recurrence sequences dominates the others.

\begin{figure}
\ifpdf
\includegraphics[scale=0.3]{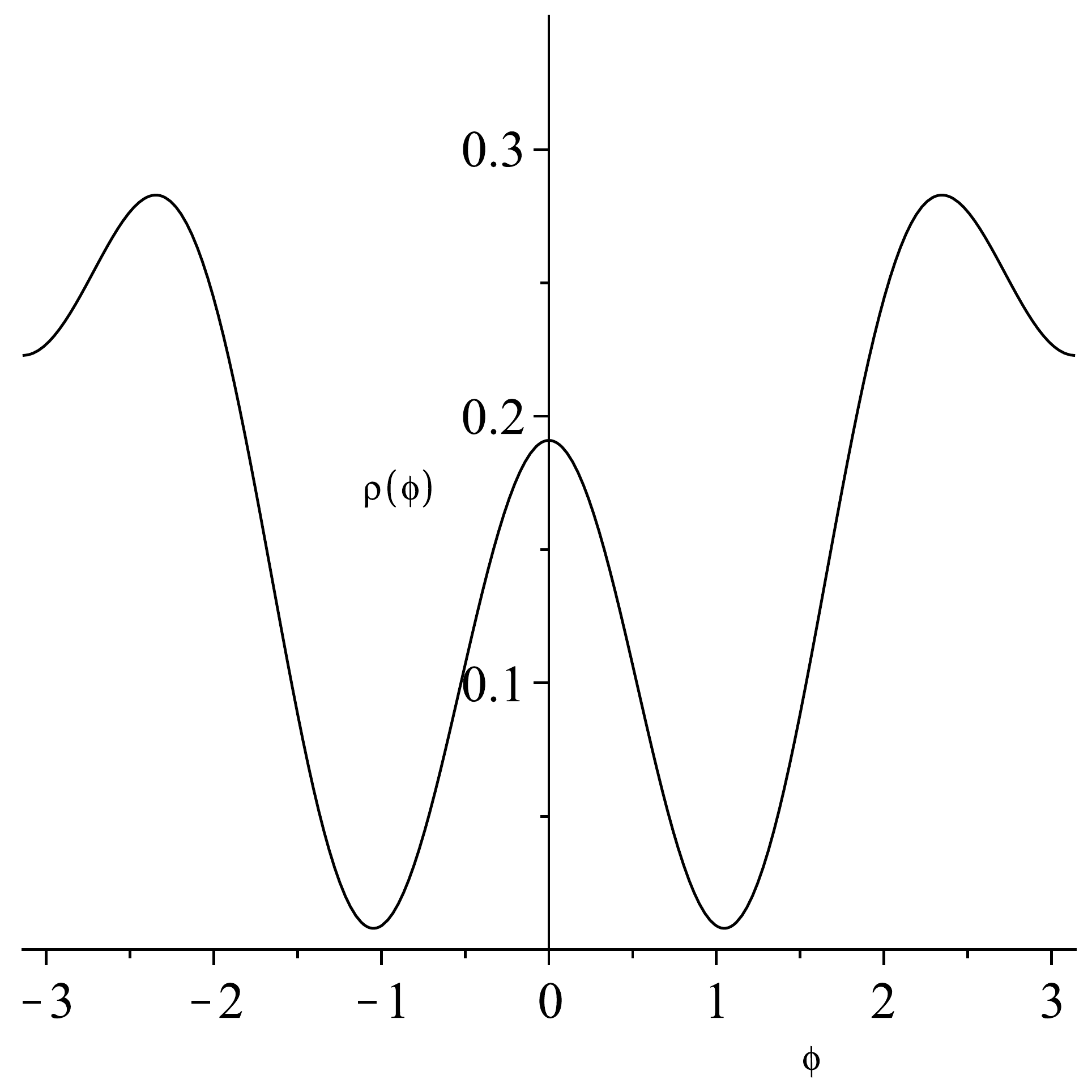}
\else
\includegraphics[scale=0.3]{f7-a.eps}
\fi
\ifpdf
\includegraphics[scale=0.5]{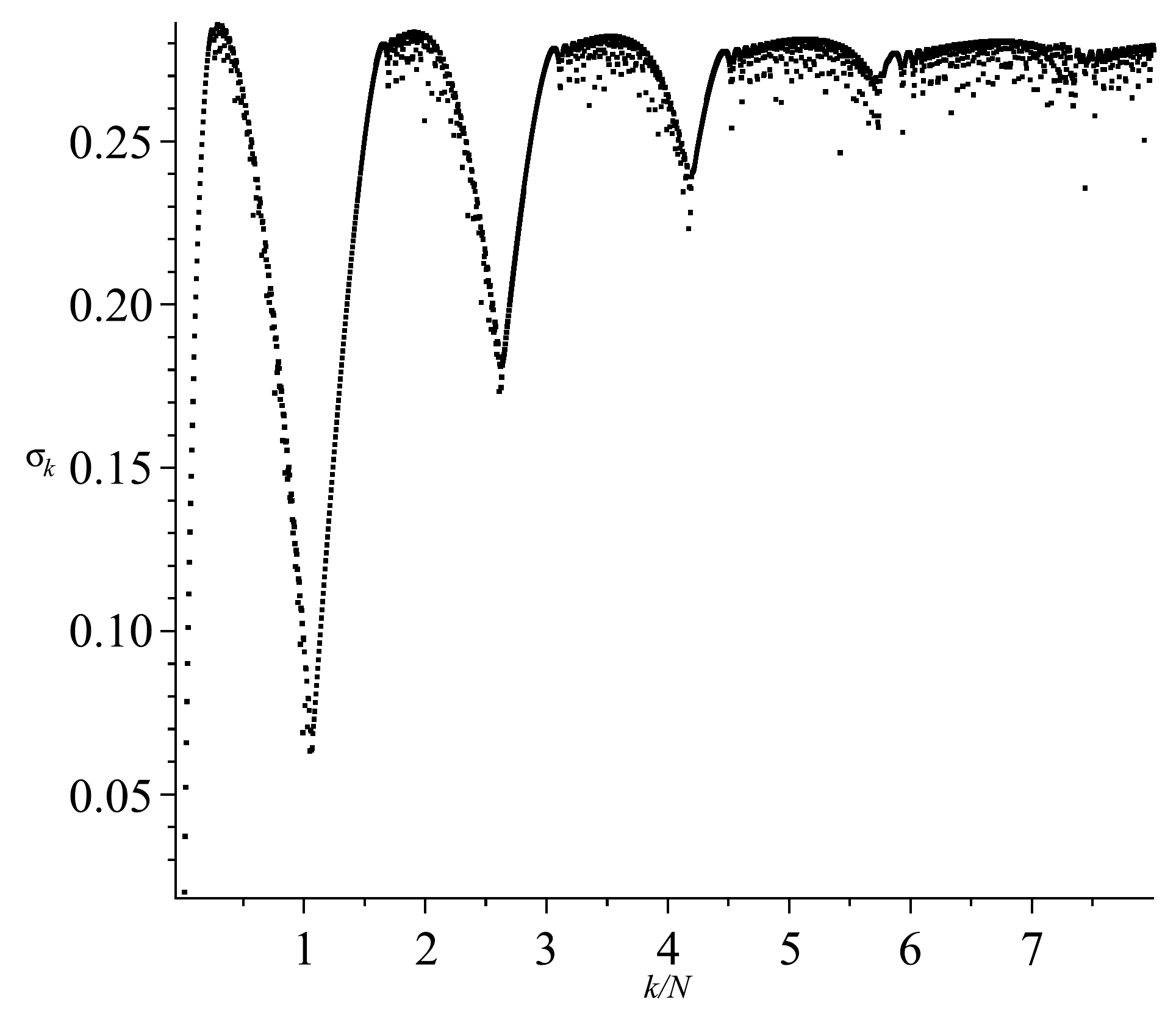}
\else
\includegraphics[scale=0.5]{f7-b.eps}
\fi
\caption{$\rho(\phi)$ given in equation (\ref{two-peak})
and the corresponding  $\sigma_k$ as a function of $k/N$,
for $N=200$.}
\label{f7}
\end{figure}

Other phenomena are possible as well.  For example, in Figure \ref{f7},
with an ungapped distribution given by
\be
\rho(\phi)={1\over 2\pi} \left ( -0.3\cos(\phi) + 0.15\cos(2\phi) +0.25\cos(3\phi)
\right )~,
\label{two-peak}
\ee
we see that the peaks can broaden and eventually merge as
they reoccur.

The recurrences as a function of $k/N$ seem to be a fairly
generic feature of $\Gamma_k$ at large $N$.  This is interesting
partly because for many theories of interest, such as ${\cal{N}}=4$ SYM,
the eigenvalue density is not known.  Therefore, while we cannot
compute $\Gamma_k$ in ${\cal{N}}=4$ SYM, we can make a conjecture that
it will exhibit some kind of recurrences.  Notice further that
$\Gamma_k$ seems also to generically experience intriguing first-order
phase transitions (kinks) as a function of $k/N$.

In the Introduction, we have already noted the qualitative difference between
the behaviour of $\langle \mathrm{Tr}_{{\cal S}_k} U \rangle$ and
$\langle \mathrm{Tr}~U^k \rangle$.  Not only is the former exponentially larger than the
latter, but it exhibits a very clean large $N$ limit which can be obtained
from the saddle point approximation.  In contrast,  $\langle \mathrm{Tr}~U^k \rangle$
cannot be obtain this way.  This suggests that if one is interested in
studying the moments of the eigenvalue distribution,
$\langle \mathrm{Tr}_{{\cal S}_k} U \rangle$  is a better object to compute than
$\langle \mathrm{Tr}~U^k \rangle$.  These two objects contain in principle
the same information, but organized in a different way.  In particular, through exponentiation,
$\langle \mathrm{Tr}_{{\cal S}_k} U \rangle$
magnifies certain features of the eigenvalue distribution $\rho$
(such as the value of $\rho_{max}$) making them easier to extract.

Since $\langle \mathrm{Tr}_{{\cal S}_k} U \rangle$  has a good large-N
limit, it is also the natural candidate for computations using the
AdS-CFT duality (the same is true of the expectation value of a  character
in the rank $k$ totally anti-symmetric tensor representation, as
was shown by similar methods in \cite{Karczmarek:2010ec}).  It would be
most interesting to find a stringy counterpart of the recurrences
conjectured in this Paper.

Finally, the potentially complex recurrence pattern means that
computing  $\langle \mathrm{Tr}_{{\cal S}_k} U \rangle$ in
finite temperature ${\cal{N}}=4$ SYM on a sphere
(for example, through the AdS-CFT duality) might teach us
about more than just the eigenvalue distribution.  If the
picture we presented in the Introduction, in which the
recurrences are due to presence of bound states of partons,
is correct, the recurrence pattern directly carries information about
the dynamics of quarks in SYM.


\section*{Acknowledgments}
This work is supported by NSERC of Canada.  GWS acknowledges
the Aspen Center for Physics,  KITP Santa Barbara, Galileo Galilei Institute
and Nordita, where parts of this work were completed. Work done at KITP is supported in part by the National
Science Foundation under Grant No.~NSFPHY05-51164 and in part by DARPA under Grant No.
HR0011-09-1-0015 and by the National Science Foundation under Grant
No. PHY05-51164.
%


\end{document}